\documentclass[12pt,preprint]{aastex}
\def\jcap{JCAP}
\def\beq{\begin{equation}}
\def\eeq{\end{equation}}
\def\ben{\begin{eqnarray}} 
\def\een{\end{eqnarray}}

\usepackage{xcolor}

\begin{document}
\title{Galaxy Spin Transition Driven by the Misalignments between the Protogalaxy Inertia and Initial Tidal Tensors}
\author{Jun-Sung Moon$^{1,2}$ and Jounghun Lee$^{1}$}
\affil{$^1$Department of Physics and Astronomy, Seoul National University, Seoul 08826, Republic of Korea
\email{cosmos.hun@gmail.com,jsmoon.astro@gmail.com}}
\affil{$^2$Research Institute of Basic Sciences, Seoul National University, Seoul 08826, Republic of Korea}
\begin{abstract}
A numerical detection of the $\tau$-driven transition of galaxy spins is presented, where $\tau$ is the degree of misalignment between the initial tidal field and protogalaxy inertia 
tensors. Analyzing the data from the IllustrisTNG 300-1 simulations, we first measure the values of $\tau$ at the protogalactic sites found by tracing the constituents of the galactic 
halos in the mass range of $10.5\le \log \left[M_{h}/(h^{-1}M_{\odot})\right] \le 13$ back to the initial stage, $z_{i}=127$. The probability density functions of $\tau$ are shown to be well 
modeled by the $\Gamma$-distributions, whose shape and scale parameters turn out to have universal values on a certain critical scale. 
Then, we investigate how the strength and tendency of the galaxy spin alignments with the principal axes of the local tidal fields depend on the initial condition, $\tau$. 
It is found that on the scale lower than the critical one, the galaxy spin transition occurs at two different thresholds from the major to intermediate and from the intermediate to minor 
principal axes of the local tidal fields, respectively. Noting that the $\tau$-dependent spin transition supersedes in strength the previously found mass-dependent, 
morphology-dependent, and radius-dependent counterparts, we suggest that $\tau$ should be the key driver of all types of the galaxy spin transition and that the present galaxy 
spins are indeed excellent fossil records of the initial condition. 
\end{abstract}
\section{Introduction}\label{sec:intro}

The galaxy spin transition is an unequivocal phenomenon, the occurrence of which was witnessed by numerous numerical studies and several observational works 
as well 
\citep[e.g.,][]{ara-etal07,hah-etal07,paz-etal08,cod-etal12,TL13,tem-etal13,tro-etal13,lib-etal13,AY14,dub-etal14,for-etal14,cod-etal15a,cod-etal15b,pah-etal16,hir-etal17,cod-etal18,gan-etal18,wan-etal18,LL20,lee-etal20,kra-etal20,wel-etal20,LM22,LM23,ML23a,ML23b,ML23c}.  
Three different types of the galaxy spin transition were reported to exist, among which the best known and widely explored type is the mass-dependent spin 
transition, which will be called the type I spin transition throughout this paper.
The low-mass (high-mass) galaxies tend to have their spins aligned with the minor (intermediate) principal axes of the local tidal fields 
\citep[e.g.,][]{ara-etal07,hah-etal07,paz-etal08,hah-etal10,lib-etal13,AY14,for-etal14,wan-etal18,lee-etal20,LL20}.

The type II is the morphology-dependent transition witnessed mainly in the observational data. The elliptical (spiral) galaxies exhibit 
perpendicular (parallel) alignments of their spins with the elongated axes of their hosting filaments \citep[e.g.,][]{tem-etal13,pah-etal16,hir-etal17,wel-etal20}.
This type II spin transition was initially thought of as a mere reflection of the type I spin transition since the ellipticals are on average more massive than the spirals 
\citep[e.g.,][]{tem-etal13,wel-etal20}. However, the latest observational analysis \citep{LM23} has revealed that even at fixed mass the galaxies with different 
morphologies located on void surfaces exhibit different alignment tendencies relative to the directions toward the void centers, implying that the type I and II spin transitions 
may not necessarily share the same origin \citep{LM23}.

The type III is the radius-dependent transition found in our prior work \citep{ML23a}. The preferred directions of the galaxy inner spins (i.e.,  
the directions of galaxy angular momentum vectors measured at more inner radii than the virial boundaries) transit from the intermediate to the major principal axes of the 
local tidal fields when the inner radii decrease from the virial boundaries down to the stellar half-mass radii, 
at which the observable stellar spins of the galaxies are usually measured in practice \citep[e.g.,][]{wel-etal20}. In filament environments, the major and intermediate principal 
axes of the local tidal fields are degenerate with each other \citep{zel70}, and thus it is difficult to distinguish between the types I and III.  In non-filamentary environments, however, 
the occurrence of the type III spin transition was found to occur regardless of galaxy mass \citep{ML23b}. 

The occurrence of the galaxy spin transitions was regarded as a counter-evidence against the standard linear tidal torque theory (LTTT), according to which 
a protogalactic region is gradually torqued by the surrounding anisotropic matter distribution till the turn-around moment to develop a tendency of its angular momentum 
vector being aligned with the intermediate principal axes of the initial tidal field \citep{dor70,whi84,CT96,lp00,lp01}.  
 The notion of the angular momentum conservation combined with the LTTT led to the unique prediction that even the present galaxies should retain the initially 
induced spin alignments with the intermediate principal axes of the local tidal fields, regardless of their masses \citep{lp00,LE07,wan-etal18}. 
This LTTT-based prediction was invalidated by the numerical discovery of the mass-dependent galaxy spin transitions 
\citep[e.g.,][]{ara-etal07,hah-etal07,paz-etal08,hah-etal10,lib-etal13,AY14,for-etal14,wan-etal18,lee-etal20,LL20}. 
The limitations of the LTTT were also noted by several works that investigated the nonlinear evolution of galaxy spin alignments \citep[e.g.,][]{lop-etal19,lop-etal21} and 
systematic effects of intrinsic galaxy alignments on the measurements of weak lensing signals \citep[e.g.,][]{shi-etal21,mai-etal23}.

It became conventional and prevalent to ascribe the failure of the original LTTT to the fact that the effect of gravitational merging on the nonlinear evolution of galaxy angular 
momentum was not properly taken into account. 
As the galaxies experience merging events, the directions of their angular momentum vectors swing relative to the surrounding large-scale structures
 \citep{dub-etal14,cod-etal15a,cod-etal15b,pah-etal16,hir-etal17,cod-etal18,gan-etal18,lop-etal19,kra-etal20,wel-etal20,lop-etal21}. 
The spin alignment tendency of high-mass galaxies should be different from that of low-mass counterparts since the former formed through more frequent 
merging events than the latter. 
This conventional and apparently plausible scenario proposed to overcome the limitation of the LTTT, however, turned out to be inconsistent with the recent 
numerical result that the strength and tendency of the halo spin alignments exhibit very weak, if existent, dependence on the latest merging epochs \citep{LM22}. 

Furthermore, it was pointed out that although the galaxy spins swing during the merging events, the tendency of their spin alignments with the local tidal fields 
would not change \citep{for-etal14,lee-etal18,wan-etal18,ML23b}. Since the orbital angular momentum of the galaxy progenitors themselves are also aligned with 
the intermediate principal axes of the local tidal fields, the spin angular momentum of the merged galaxies transferred from the orbital angular momentum should retain the 
same alignments tendency as the original LTTT predicted, although the scales would change. 
Most importantly, this conventional scenario based on the build-up of mass through merging process 
cannot easily accommodate the aforementioned type II and III spin transitions, which were found to occur even at fixed mass \citep{lib-etal13, TL13, ML23a,ML23b, LM23}.

In our prior work \citep{ML23b}, a new scenario, dubbed the {\it density parity model}, was proposed for the radius-dependent spin transition, according to which the 
competition between the tidal compression and inner tension determines which principal axes of the tidal fields are aligned with the galaxy inner spins. 
Based on this model, a purely analytic formula for the transition threshold radii was derived and numerically confirmed to be valid in describing when the radius-dependent 
spin transition occurs and how the threshold radius evolves with redshift and varies with galaxy mass. 
Nevertheless, the density parity model is still incapable of coherently explaining the type I and II transitions of galaxy {\it virial} spins, i.e., the directions of galaxy angular 
momentum vectors measured at the virial boundaries \citep{ML23b}.

Very recently,  however, an interesting clue is provided by \citet{ML23d} who have found that the dimensionless spin parameters of the present galaxies ($\lambda$) have 
an almost perfectly linear correlation with the degree of misalignments between the principal axes of the protogalaxy inertia and initial tidal tensors ($\tau$). 
If the magnitudes of the galaxy angular momentum vectors quantified by $\lambda$ \citep{bul-etal01} are that strongly correlated with the initial condition $\tau$, 
then it is quite natural to expect that their directions should be also correlated with $\tau$.  
Here, we are going to investigate if and how strongly the initial condition $\tau$ affects the galaxy spin alignments with the principal axes of the local tidal fields, 
in the hope that the aforementioned three types of galaxy spin transitions might be coherently and coherently understood if the impact of $\tau$ 
on the galaxy spin orientations is properly incorporated into the LTTT. 
The organization of this paper is as follows. In Section \ref{sec:review} we critically review the LTTT,  examining its limitations. 
In Section \ref{sec:test}, we describe the numerical data and the procedures through which the effect of $\tau$ on the galaxy spin alignments is disclosed. 
In Section \ref{sec:con}, we summarize the primary results and discuss their implications. 

\section{A Critical Review of the LTTT}\label{sec:review}

The original LTTT states that a protogalaxy at redshift, $z_{i}\gg 1$, gradually acquires the angular momentum, ${\bf J}(z_{i})$, at first order 
until the turn-around moment, as it experiences the torque force exerted by the initial tidal field, ${\bf T}(z_{i})$, on the protogalaxy mass scale \citep{dor70,whi84}. 
In the principal frame of ${\bf T}(z_{i})$ spanned by its three orthonormal eigenvectors, $\{\hat{\bf t}_{1}(z_{i}),\ \hat{\bf t}_{2}(z_{i}),\ \hat{\bf t}_{3}(z_{i})\}$, 
three components of the protogalaxy angular momentum vector, ${\bf J}(z_{i})$,  can be expressed as \citep{lp00}
\begin{eqnarray}
\label{eqn:lttt1}
J_{1}(z_{i}) &\propto& I_{23}(z_{i})\left[u_{2}(z_{i})-u_{3}(z_{i})\right]\, , \\
\label{eqn:lttt2}
J_{2}(z_{i}) &\propto& I_{31}(z_{i})\left[u_{3}(z_{i})-u_{1}(z_{i})\right]\, ,  \\
\label{eqn:lttt3}
J_{3}(z_{i}) &\propto& I_{12}(z_{i})\left[u_{1}(z_{i})-u_{2}(z_{i})\right]\, ,
\end{eqnarray}
where $\{u_{1}(z_{i}),\ u_{2}(z_{i}),\ u_{3}(z_{i})\}$ are three eigenvalues of ${\bf T}(z_{i})$ in a decreasing order, corresponding to 
$\{\hat{\bf t}_{1}(z_{i}),\ \hat{\bf t}_{2}(z_{i}),\ \hat{\bf t}_{3}(z_{i})\}$, respectively, and  $\{I_{12}(z_{i}),\ I_{23}(z_{i}),\ I_{31}(z_{i})\}$, are three off-diagonal components 
of the protogalaxy inertia tensor, ${\bf I}(z_{i})$, in the principal frame of ${\bf T}(z_{i})$. 

Equations (\ref{eqn:lttt1})--(\ref{eqn:lttt3}) imply that the misalignment between the principal axes of ${\bf I}(z_{i})$ and ${\bf T}(z_{i})$ is the necessary condition 
for the first order generation of ${\bf J}(z_{i})$. \citet{ML23d} suggested that the degree of the misalignments between the principal axes of 
${\bf I}(z_{i})$ and ${\bf T}(z_{i})$ be quantified as
\begin{equation}
\label{eqn:tau}
\tau \equiv 
\left[\frac{{I}^{2}_{12}(z_i) + {I}^{2}_{23}(z_i) + {I}^{2}_{31}(z_i)}
{{I}^{2}_{11}(z_i)+{I}^{2}_{22}(z_i)+{I}^{2}_{33}(z_i)}\right]^{1/2}\, .
\end{equation}
where $\{I_{11}(z_{i}),\ I_{22}(z_{i}),\ I_{33}(z_{i})\}$, are three diagonal components of ${\bf I}(z_{i})$. 
The more the principal axes of ${\bf I}(z_i)$ and ${\bf T}(z_i)$ are misaligned with one another, the larger value the initial condition $\tau$ will have. 
The Zel'dovich approximation \citep{zel70} on which the LTTT was based presumes a perfect anti-alignments between the principal axes of ${\bf I}(z_{i})$ and 
${\bf T}(z_{i})$, that is, $\tau=0$. In reality, however, it has been shown by $N$-body experiments that although 
the principal axes of the two tensors are on average quite strongly aligned with one another, $\tau$ would deviate from $0$, 
rendering ${\bf J}(z_{i})$ to be generated at first order \citep{lp00,lp01,por-etal02}. 

Noting in Equations (\ref{eqn:lttt1})--(\ref{eqn:lttt3}) that the magnitude of the difference between  $u_{3}(z_{i})$ and $u_{1}(z_{i})$ should be larger than those between 
$u_{2}(z_{i})$ and $u_{1}(z_{i})$ and between $u_{2}(z_{i})$ and $u_{1}(z_{i})$, \citet{lp00} predicted that ${\bf J}(z_{i})$ will generically develop a tendency of being 
preferentially aligned with $\hat{\bf t}_{2}(z_{i})$.  Strictly speaking, this claim is contingent upon the validity of the following two approximations: 
\begin{eqnarray}
\label{eqn:approx1}
&&\langle\vert I_{ab}\left( u_{a} - u_{b}\right)\vert\rangle \approx
\langle\vert I_{ab}\vert\rangle\langle\vert u_{a} - u_{b}\vert\rangle\, ,\,\, (a, b)\in\{(1,2),(2,3),(3,1)\}\, ,\\
\label{eqn:approx2}
&&\langle\vert I_{12}\vert\rangle \approx  \langle\vert I_{23}\vert\rangle \approx  \langle\vert I_{31}\vert\rangle.
\end{eqnarray}
The very existence of strong mean anti-alignments between the principal axes of ${\bf I}(z_{i})$ and ${\bf T}(z_{i})$, however, invalidates the first approximation given in 
Equation~(\ref{eqn:approx1}).  To address this issue, \citet{lp00} suggested that the strong alignments between the two tensors should be treated as approximation errors, 
which play the role of randomizing the direction of ${\bf J}(z_{i})$ with respect to the principal axes of ${\bf T}(z_{i})$. Then, they proposed the following empirical formula to 
effectively describe the expected alignments between $\hat{\bf j}(z_{i})\equiv {\bf J}(z_{i})/|{\bf J}(z_{i})|$ and $\hat{\bf t}_{2}(z_{i})$ in the presence of strong mean anti-alignments 
between the principal axes of ${\bf I}(z_{i})$ and ${\bf T}(z_{i})$: 
\begin{equation}
\label{eqn:lp00}
\langle\hat{j}_{a}\hat{j}_{b}\rangle = \left(\frac{1}{3}+\frac{1}{5}\kappa\right)\delta_{ab} - \frac{3}{5}\kappa\,(\hat{T}_{ad}\hat{T}_{db})\, , \quad a,b\in\{1,\ 2,\ 3\}\, ,
\end{equation}
where $\delta_{ab}$ is the Kroneker-delta symbol, $\hat{\bf T}(z_{i})$ is the unit traceless version of ${\bf T}(z_{i})$,   and $\kappa$ is a free parameter in 
the range of $[0,1]$.  The expected value of $\kappa$ depends on how strongly the principal axes of ${\bf I}(z_{i})$ are anti-aligned with those of ${\bf T}(z_{i})$.
A perfect anti-alignment would completely randomize $\hat{\bf j}(z_i)$ relative to the principal axes of ${\bf T}(z_{i})$, resulting in $\kappa=0$.  
A perfect {\it misalignment} would maximize the strength of the $\hat{\bf j}(z_{i})$--$\hat{\bf t}_{2}(z_{i})$ alignments, leading to $\kappa=1$. 
Meanwhile, a strong but imperfect alignment between ${\bf I}(z_{i})$ and ${\bf T}(z_{i})$ as found in $N$-body simulations would yield a low but 
non-negligible value of $\kappa$. 

\citet{lp00} suggested further that Equation (\ref{eqn:lp00}) should be able to effectively describe the tidally induced alignments of the galaxy spins not only 
at the protogalactic stages but even at much later epochs, $z\ll z_{i}$,  by treating $\kappa$ as a redshift dependent parameter, since the angular momentum is 
a conserved quantity. This idea of \citet{lp00} was supported by several $N$-body experiments and a couple of observational works \citep[e.g.,][]{nav-etal04,LE07,zha-etal15} 
in the high-mass section. 
Yet, the results from recent high-resolution simulations invalidated Equation (\ref{eqn:lp00}) in the highly nonlinear regions on the dwarf-galactic scales. 
No matter what value $\kappa$ has, Equation (\ref{eqn:lp00}) fails to describe the $\hat{\bf j}$--${\bf t}_{3}$ alignments found from the dwarf galactic halos at $z\sim 0$ 
\citep[e.g.,][]{ara-etal07,lib-etal13} as well as the $\hat{\bf j}$--${\bf t}_{1}$ alignments found at the inner radii of galactic halos $r$ much smaller than their virial radii 
$r_{\rm vir}$ \citep{ML23a}. 

Several attempts were made in the direction of modifying and improving the LTTT to account for the nonlinear evolution of the galaxy spin alignments. 
For example, \citet{viv-etal02} took into account the effect of hierarchical merging on the evolution of galaxy angular momentum vectors.
\citet{lp08} added a new term, $\epsilon_{nl}\hat{T}_{ab}$,  to Equation (\ref{eqn:lp00}) with an extra free parameter $\epsilon_{nl}$ 
to describe the deviation of $\hat{\bf j}$ from the LTTT prediction in the nonlinear regime \citep[see also ][]{lop-etal19,lop-etal21}. 
\citet{lib-etal13} suggested that the generation of vorticity in the nonlinear regime should affect the galaxy angular momentums, causing 
the failure of the original LTTT. 
\citet{cod-etal15b} incorporated the effect of anisotropic environments into the LTTT to explain the mass-dependent transition of galaxy 
spins and to describe the redshift evolution of the transition threshold mass. 
Notwithstanding, none of them are able to provide a coherent and consistent explanation for the occurrences of the three types of galaxy spin transitions 
and complexity involved in the transition thresholds, described in Section \ref{sec:intro}. 

In the current work, we tackle this issue from a different direction. Instead of trying to figure out what deviates $\hat{\bf j}$ from the LTTT prediction in the 
nonlinear regime, we reexamine from the scratch whether or not the LTTT actually predicts the mass independent $\hat{\bf j}$--$\hat{\bf t}_{2}$ alignments, 
suspecting that the second approximation of Equation (\ref{eqn:approx2}), used to derive Equation (\ref{eqn:lp00}), may have a limited validity.  
This approximation is valid only provided that $\tau$ is low enough for the three off-diagonal components, $I_{12}$, $I_{23}$, and $I_{31}$ 
to make negligible contribution to $\hat{\bf j}(z_i)$.  Although the numerical experiments confirmed the low average value of $\tau\lesssim 0.2$ \citep{lp00,por-etal02}, 
its scatter could be large if the probability distribution of $\tau$ has a broad shape. 
A large value of $\tau$ translates into large values of $I_{12}$, $I_{23}$, and $I_{31}$ in Equations (\ref{eqn:lttt1})--(\ref{eqn:lttt3}), the differences among 
which could play the decisive role for the determination of the orientations of protogalaxy angular momentum vectors rather than the differences in the eigenvalues 
of the initial tidal field. 
This critical review of the LTTT motivates a speculation that its generic prediction may not necessarily be the universal $\hat{\bf j}$--$\hat{\bf t}_{2}$ 
alignments. Rather,  it might be possible to explain the spin transitions in the frame of the LTTT if the differences in the values of $\tau$ among the protogalaxies 
are properly taken into account. In Section \ref{sec:test}, a numerical test of this speculation will be carried out by resorting to a high-resolution hydrodynamical simulation. 
 
\section{A Numerical Test}\label{sec:test}

To numerically explore the effect of the initial condition $\tau$ on the orientations of galaxy angular momentum vectors relative to the principal axes of the local tidal fields, 
we use a dataset from the TNG300-1 cosmological hydrodynamical simulation conducted 
as a part of the IllustrisTNG project \citep{tngintro1, tngintro2, tngintro3, tngintro4, tngintro5, illustris19}  for the Planck universe \citep{planck16}. 
The simulation volume ($V_{\rm box}$), numbers ($N_{\rm dm}$, $N_{\rm b}$) and individual masses ($m_{\rm dm}$, $M_{\rm b}$) of dark matter (DM) 
and baryonic particles, respectively, and starting redshift ($z_{i}$) of the TNG300-1 simulation are as follows: $V_{\rm box}/(h^{-3}{\rm Mpc}^{3})=205^{3}$, 
$N_{\rm dm}=2500^{3}$, $N_{\rm b}=2500^{3}$, $m_{\rm dm}/(10^{7}M_{\odot})=5.9$, $m_{\rm b}/(10^{7}M_{\odot})=1.1$ and $z_{i}=127$. 
Applying the same algorithm as in \citet{LM22} to the particle snapshot at $z_{i}=127$, 
we construct the initial tidal field on $512^{3}$ grids, ${\bf T}(z_{i})$, convolved with a Gaussian kernel on the comoving scale of $R_{f}$. 
Repeating the same procedure but at some later epoch $z(\ll z_{i})$, we also construct the local tidal field, ${\bf T}(z)$. 
The computational steps taken to construct the tidal fields can be summarized as: 
(i)  an application of the cloud-in-cell method to the spatial distribution of DM and gas particles to obtain a raw density contrast field, $\delta_{{\rm raw}}$, on 
each grid point ${\bf x}$ in real space; 
(ii) a Fourier-transformation of $\delta_{{\rm raw}}({\bf x})$ into $\tilde{\delta}_{{\rm raw}}({\bf k})$ with wave vector ${\bf k}=(k_{a})$; 
(iii) an inverse Fourier-transformation of $\tilde{T}_{ab}({\bf k})\equiv k_{a}k_{b}\tilde{\delta}_{{\rm raw}}({\bf k})\exp(-\vert{\bf k}\vert^{2}R^{2}_{f}/2)/\vert{\bf k}\vert^{2}$ 
into $T_{ab}({\bf x})$. 

From the TNG 300-1 catalogs of distinct halos and their subhalos identified at redshift $z$ via the friends-of-friends (FoF) and SUBFIND 
algorithms \citep{subfind}, respectively, we first cull only those galactic subhalos whose total masses, $M_{\rm h}$, fall in the range of 
$10.5\le m_{h}\le 13$ with $m_{h}\equiv \log\left[M_{\rm h}/(h^{-1}\,M_{\odot})\right]$, consisting of $300$ or more particles \citep{bet-etal07}.  
We consider only the central subhalos, i.e., the most massive subhalos in each FoF halo, to avoid complication introduced by strong environmental effects inside 
massive groups of galaxies.
From here on, the selected subhalos and directions of their angular momentum vectors will be referred to as the galaxies and the galaxy spins, respectively. 
Tracing back the component DM particles of each selected galaxy back to $z_{i}$, we determine its Lagrangian center of mass, $\bar{\bf q}$, 
as its protogalactic site and the six independent components of the protogalaxy inertia tensor, $\widehat{\bf I}(z_{i})$, as 
\begin{equation}
\widehat{I}_{ab}({\bf q})= \sum_{\mu=1}^{n_{p}}m_{\mu}\left(q_{a,\mu}-\bar{q}_{a})(q_{b,\mu}-\bar{q}_{b}\right)\, ,
\end{equation} 
where ${\bf q}_{\mu}=(q_{a,\mu})$ is the initial position of the $\mu$th particle, and $n_{p}$ is the total number of the component particles. 

At the grid point where $\bar{\bf q}$ falls,  we diagonalizing the initial tidal tensor, ${\bf T}(z_i)$, to determine its three orthonormal eigenvectors, 
$\{\hat{\bf t}_{a}(z_i)\}_{a=1}^{3}$,  corresponding to its largest, second largest and smallest eigenvalues
as the major, intermediate and minor principal directions.   
Creating a rotation matrix, ${\bf R}(z_i)$, whose three columns equal $\{\hat{\bf t}_{a}(z_i)\}_{a=1}^{3}$, we perform a similarity transformation 
of $\widehat{\bf I}(z_{i})$ to evaluate its new form, ${\bf I}(z_{i})$,  {\it in the principal frame} of ${\bf T}(z_i)$ as 
${\bf I}(z_{i})\equiv {\bf R}^{-1}(z_i)\cdot \widehat{\bf I}(z_{i})\cdot {\bf R}(z_i)$. 
At each protogalactic site, the degree of misalignment, $\tau$, between the principal axes of ${\bf I}(z_i)$ and ${\bf T}(z_i)$ is computed by 
Equation (\ref{eqn:tau}) \citep{ML23d} 

Binning the range of $\tau$ into multiple short intervals of equal size $\Delta\,\tau$ and counting the number, $n_{g}(\tau)$,  of the galaxies in 
each $\tau$-bin, we compute the probability density function as $p(\tau)\equiv n_{g}/(N_{g}\,\Delta\tau)$, where $N_{g}$ is the total number of galaxies.  
Performing the bootstrap error analysis, we compute 
the errors in $p(\tau)$ at each $\tau$-bin as one standard deviation scatter among $10,000$ different bootstrap resamples each of which is created by extracting 
the same  number of the galaxies but with repetition allowed. In our prior work \citep{ML23d},  the galaxy spin parameters, $\lambda$, were shown to have an almost 
perfect linear correlation with $\tau$ and that the probability density function of $\lambda$ is better approximated by the $\Gamma$-distribution than by the conventional 
log-normal counterpart \citep{bul-etal01}.  In light of these prior results, we investigate here whether or not $p(\tau)$ can also be analytically described by the following 
$\Gamma$-distribution characterized by the scale and shape parameters, $k$ and $\theta$, respectively: 
\begin{equation}
\label{eqn:gam}
p(\tau;k,\theta) = \frac{1}{\Gamma(k)\theta^{k}}\tau^{k-1}\exp\left(-\frac{\tau}{\theta}\right)\, .
\end{equation}

With the help of the $\chi^{2}$-minimization, we adjust Equation (\ref{eqn:gam}) to the numerically computed probability density, $p(\tau)$, to find 
the best-fit values of $k$ and $\theta$ (see Table \ref{tab:bestfit1}). 
Figure \ref{fig:pro_tau} shows the numerically computed $p(\tau)$ with the jackknife errors for three different cases of $r_{f}\equiv R_{f}/(h^{-1}{\rm Mpc})=0.5, 1$ and $2$ 
(red, blue and green filled circles, respectively) in three different ranges of $m_{h}$ at $z=0$ and $1$ (left and right panels, respectively). 
It also shows the corresponding best-fit $\Gamma$-distributions (red, blue and green solid lines), revealing excellent agreements between the numerically computed $p(\tau)$ and 
Equation (\ref{eqn:gam}) for all of the considered cases of $m_{h}$,  $r_{f}$ and $z$.  Noting a somewhat systematic change of $p(\tau)$ with $r_{f}$ and $m_{h}$ at each 
redshift, we suspect that $p(\tau)$ may have a universal shape, regardless of $z$ and $m_{h}$, if $r_{f}$ is properly chosen in each $m_{h}$-range.  
Varying $r_{f}$, repeatedly computing $p(\tau)$ and finding the best-fit parameters (see Table \ref{tab:bestfit2}), we discover that $p(\tau)$ indeed 
has an universal shape on a mass-dependent scale of $r_{f}=8\bar{r}_{\rm vir}$ where $\bar{r}_{\rm vir}$ is the mean virial radius in unit of $h^{-1}$Mpc averaged over the galaxies 
in a given $m_{h}$-range.  
In  each panel of Figure \ref{fig:pro_tau} is also plotted this universal distribution, $p(\tau)$, for the case of  $r_{f}=8\bar{r}_{\rm vir}$ (black filled circles and solid line).  
Note that $p(\tau)$ tends to spread out more broadly as $r_{f}$ deviates further from the critical smoothing scale, $8\bar{r}_{\rm vir}$, required for its universality.  

Binning the $\tau$ values of the galaxies in each $m_{h}$-range, we evaluate three ensemble averages, $\{\langle\vert\hat{\bf j}(z)\cdot\hat{\bf t}_{a}(z)\vert\rangle\}_{a=1}^{3}$, 
separately over each $\tau$-bin. Here, the galaxy spins are computed as 
$\hat{\bf j}\equiv {\bf J}/\vert{\bf J}\vert$ with ${\bf J} = \sum_{\mu=1}^{n_{p}}{m_{\mu}}({\bf x}_{\mu}-\bar{\bf x})\times ({\bf v}_{\mu}-\bar{\bf v})$
where $m_{\mu}, {\bf x}_{\mu}$ and ${\bf v}_{\mu}\}$ denote the mass, comoving position and peculiar velocity of the $\mu$th component particle, respectively. 
The jackknife method is also employed to compute the errors in $\{\langle\vert\hat{\bf j}(z)\cdot\hat{\bf t}_{a}(z)\vert\rangle\}_{a=1}^{3}$.
The left three panels of Figures~\ref{fig:align_z0_rf0.5}-\ref{fig:align_z1_rf0.5} plot $\{\langle\vert\hat{\bf j}\cdot\hat{\bf t}_{a}\vert\rangle\}_{a=1}^{3}$ with the jackknife errors 
versus the median values of $\tau$ in the three $m_{h}$-ranges for the case of $r_{f}=0.5$ at $z=0$ and $1$, respectively.  
The left three panels of Figures~\ref{fig:align_z0_rf1}--\ref{fig:align_z1_rf1} plot the same mean spin alignments but for the case of $r_{f}=1$ and compare them with the results 
for the case of $r_{f}=0.5$ (thin lines). As can be seen, the mean spin alignments change their tendencies and strengths with $\tau$ for all of the considered cases of $m_{h}$ and $z$. 
Especially, the $\hat{\bf j}$--$\hat{\bf t}_{3}$ alignment shows a very strong dependence on $\tau$, consistently increasing as $\tau$ increases for all of the considered cases. 

From the results shown in the left panels of Figures \ref{fig:align_z0_rf0.5}--\ref{fig:align_z1_rf1}, one can say that the principal axis of ${\bf T}(z)$ aligned with $\hat{\bf j}(z)$ 
indeed transit in a very $\tau$-dependent way. 
To rigorously determine the transition thresholds, we employ the threshold-finding scheme based on the Kolmogorov–Smirnov (KS) test \citep{LL20}.  
In each $\tau$-bin, we determine three probability density functions, $\{p(\vert\hat{\bf j}\cdot\hat{\bf t}_{a}\vert)\}_{a=1}^{3}$, and three cumulative probability distributions, 
$\{P(\ge\vert\hat{\bf j}\cdot\hat{\bf t}_{a}\vert)\}_{a=1}^{3}$, as well. Then, we perform the KS test of a null hypothesis, $H_{0}$, set up as 
$p(\vert\hat{\bf j}\cdot\hat{\bf t}_{2}\vert)= p(\vert\hat{\bf j}\cdot\hat{\bf t}_{3}\vert)$ (i.e., statistical equivalence). 
The type I transition zone,  $\tau_{c1}$, is located by two criteria: (i) $H_{0}$ is rejected at the confidence level lower than $99.9\%$ in the $\tau_{c1}$-bin; 
(ii) $H_{0}$ is rejected at the confidence level higher than $99.9\%$ in the $\tau$-bins adjacent to $\tau_{c1}$. 
 The rejection of $H_{0}$ at the confidence level higher than $99.9\%$ corresponds to $\tilde{D}_{\rm max}\equiv \sqrt{N_{g}/2}D_{\rm max}\ge1.949$, where 
$D_{\rm max}$ denotes the maximum distance between $P(\le\vert\hat{\bf j}\cdot\hat{\bf t}_{2}\vert)$  and $P(\le\vert\hat{\bf j}\cdot\hat{\bf t}_{3}\vert)$. 

The type II and III transition zones, $\tau_{c2}$ and $\tau_{c3}$, are also located through the same steps but with $H_{0}$ replaced by $p(\vert\hat{\bf j}\cdot\hat{\bf t}_{1}\vert)= 0.5$ 
and by $p(\vert\hat{\bf j}\cdot\hat{\bf t}_{1}\vert)= p(\vert\hat{\bf j}\cdot\hat{\bf t}_{2}\vert)$, respectively. \footnote{For the one-sample KS test used to determine $\tau_{c2}$, 
$\tilde{D}_{\rm max}\equiv \sqrt{N_{g}}D_{\rm max}$.} In the right panels of Figures \ref{fig:align_z0_rf0.5}--\ref{fig:align_z1_rf1}, we show $\tilde{D}_{\rm max}$ computed to locate 
$\tau_{c1}$, $\tau_{c2}$, and $\tau_{c3}$ (filled blue squares, red triangles, and green circles, respectively) as a function of $\tau$ for each case, and compare them with the critical 
value of $1.949$ (horizontal black dashed lines). Provided that $\tilde{D}_{\rm max}$ has a local minimum below the horizontal dashed lines, then the $\tau$-bins 
with $\tilde{D}_{\rm max}<1.949$ correspond to the transition zones (see Table \ref{tab:bestfit1}).  The thin lines in the right of Figures \ref{fig:align_z1_rf0.5}-\ref{fig:align_z1_rf1} 
shown for comparison correspond to $\tilde{D}_{\rm max}(\tau)$ for the case of $z=0$ and $r_{f}=0.5$. 

The type I spin transition between $\hat{\bf t}_{2}$ and $\hat{\bf t}_{3}$ seems to occur for all of the considered cases of $m_{h}$, $r_{f}$ and $z$.  
It is worth emphasizing here the difference in the signal strength and its variation with $m_{h}$ between the conventional and current approaches to the spin transition phenomena. 
In the conventional approach where the mean spin alignments, $\{\langle\vert\hat{\bf j}\cdot\hat{\bf t}_{a}\vert\rangle\}_{a=1}^{3}$, were expressed as a function of $m_{h}$, 
the lowest mass-range of $10.5\le m_{h}< 11$ yielded $\langle\hat{\bf j}\cdot\hat{\bf t}_{2}\rangle\approx 0$ and $\langle\vert\hat{\bf j}\cdot\hat{\bf t}_{3}\vert\rangle\lesssim 0.51$, 
while the highest mass range of $11.5\le m_{h}< 13$ produced $\langle\hat{\bf j}\cdot\hat{\bf t}_{3}\rangle\approx 0$ and 
$\langle\vert\hat{\bf j}\cdot\hat{\bf t}_{2}\vert\rangle\sim 0.52$ \citep[e.g., see Figure 4 in][]{ML23a}. 
In contrast, the current study which views $\{\langle\vert\hat{\bf j}\cdot\hat{\bf t}_{a}\vert\rangle\}_{a=1}^{3}$ as a function of $\tau$ finds 
clear and stronger signals of the $\hat{\bf j}$--$\hat{\bf t}_{2}$ and $\hat{\bf j}$--$\hat{\bf t}_{3}$ alignments in all of the three $m_{h}$-ranges. 
This result implies that the key driver of the type I spin transition may be the initial condition $\tau$ rather than the build-up of galaxy mass via hierarchical merging.

As can be seen in Figures \ref{fig:align_z0_rf0.5}--\ref{fig:align_z1_rf1}, the $\tau$-dependent spin transition between $\perp\hat{\bf t}_{1}$ and $\parallel\hat{\bf t}_{1}$ 
can be witnessed only provided that ${\bf T}$ is smoothed on the scale smaller than $8\bar{r}_{\rm vir}$ required for the universality of $p(\tau)$.  Given that $\tau$ was found 
to have strong correlations with the ages, colors, metallicities, and spin parameters of the galaxies \citep{ML23d},  which have been known to be the 
powerful indicator of galaxy morphology \citep[e.g.,][]{oh-etal13,jia-etal19,rod-etal22},  this result nicely explains why the observed spiral galaxies with different Hubble types 
exhibit different alignment tendencies relative to the directions toward to the void centers even at fixed mass \citep{ML23d}. In other words, this result also supports the notion 
that $\tau$ is the main driver of the type II spin transition. 

The results shown in Figures \ref{fig:align_z0_rf0.5}--\ref{fig:align_z1_rf1} and Table \ref{tab:bestfit1} also indicate that the $\tau$-dependent type III spin transition between 
$\hat{\bf t}_{1}$ and $\hat{\bf t}_{2}$ requires a more stringent condition of $r_{f}< 4\bar{r}_{\rm vir}$, since its occurrence is witnessed only for the case of $r_{f}=0.5$ 
and $11.5\le m_{h}\le 13$, where the stringent condition is met. 
Recall the previous numerical result that the spin transition from $\hat{\bf t}_{2}$ to $\hat{\bf t}_{1}$ was detected only at the innermost radii when the 
mean spin alignments were expressed as a function of radial distances from the galaxy centers \citep{ML23a}. 
In the current work, however,  the signals of the type III spin transition from $\hat{\bf t}_{2}$ to $\hat{\bf t}_{1}$ is detected even at virial radii, which implies that the key driver 
of the type III spin transition may be the initial condition $\tau$ rather than the differences in radial distances and any nonlinear effect inside the galaxies. 

Repeating the whole calculations but for the case of $r_{f}=8\bar{r}_{\rm vir}$, we examine how the spin alignment tendency and transition threshold change 
when ${\bf T}(z)$ is smoothed on the critical scale, the results of which are shown in Figures \ref{fig:align_z0_rfc}--\ref{fig:align_z1_rfc} and Table \ref{tab:bestfit2}. 
As can be seen, only the type I but no type II and III spin transitions are found to occur on this scale, which supports the aforementioned claim that the alignments of 
$\hat{\bf j}$ with $\hat{\bf t}_{1}$ can be witnessed only if $r_{f}$ is smaller than the critical smoothing scale.  We now suggest that the critical scale $r_{f}=8\bar{r}_{\rm vir}$, on which 
$p(\tau)$ takes on the universal form, should be regarded as the forbidden scale beyond which $\hat{\bf j}$ shows no tendency of being aligned with $\hat{\bf t}_{1}$, 
regardless of $m_{h}$ and $z$. 
Varying the lower cutoff of $m_{h}$ in the range of $[10.4, 10.6]$, we confirm that the final results are quite robust against the choice of the lower mass cut-off.  
We do not consider a lower cutoff $m_{h}\le 10.4$ for which the critical scale of $8\bar{r}_{\rm vir}$ drops below $0.4\,h^{-1}\,$Mpc. Given the particle resolution of 
the TNG300-1 simulations, the tidal fields smoothed on the scales lower than $r_{f}=0.4$ are likely to be quite noisy even on a larger number of grid points. 

\section{Summary and Discussion}\label{sec:con}

In the classical approach based on the LTTT to the origin of galaxy angular momentum, two presumptions were implicitly made: 
First, the degree of the alignment between the protogalaxy inertia and initial tidal tensors ($\tau$) would be low but non-negligibly high enough for a galaxy to acquire its 
angular momentum at first order \citep{dor70,whi84,CT96}. Second,  the initial condition $\tau$ will have no effect on the tendency and strength of the expected alignments 
between the present galaxy spins and the intermediate principal axes of the local tidal field \citep{lp00,lp01,por-etal02}. 
However, in wake of the recent numerical discovery that the initial condition $\tau$ has a strong effect on such galaxy properties as the dimensionless spin parameter, 
stellar-to-total mass ratio, morphology, age, metallicity, and color \citep{ML23d},  we have rigorously tested these two presumptions against the TNG300-1 simulations, and disproved
them by revealing that $\tau$ is in fact the key driver of the three types of galaxy spin transitions. 

The initial tidal fields smoothed by a Gaussian kernel on three different scales of $r_{f}\equiv R_{f}/(h^{-1}\,{\rm Mpc})=$ 0.5, 1 and 2, have been constructed 
from the TNG300-1 initial condition at $z_{i}=127$.  For each of the TNG galaxies in the total mass range of $10.5\le m_{h}\equiv\log\left[M_{h}/(h^{-1}\,M_{\odot})\right]<13$ 
identified at $z=0$ and $1$, its protogalactic inertia tensor has been determined from its constituent DM particles traced back to $z_{i}$.  
For each case of $m_{h}$, $z$ and $r_{f}$,  two tasks have been undertaken in the current work. First, from the values of $\tau$ measured for the TNG 
galaxies by Equation (\ref{eqn:tau}),   the probability density function, $p(\tau)$, has been determined and fitted to the analytic $\Gamma$-distribution given in 
Equation (\ref{eqn:gam}).  Second, with the local tidal field constructed from the TNG particle snapshot at each $z$, the mean alignments of its 
three principal axes ($\{\hat{\bf t}_{a}\}_{a=1}^{3}$) with the galaxy spin ($\hat{\bf j}$) has been computed as a function of $\tau$. 
The primary outcomes of these tasks and their implications are summarized as follows.

\begin{figure}[ht]
\centering
\includegraphics[height=16cm,width=15cm]{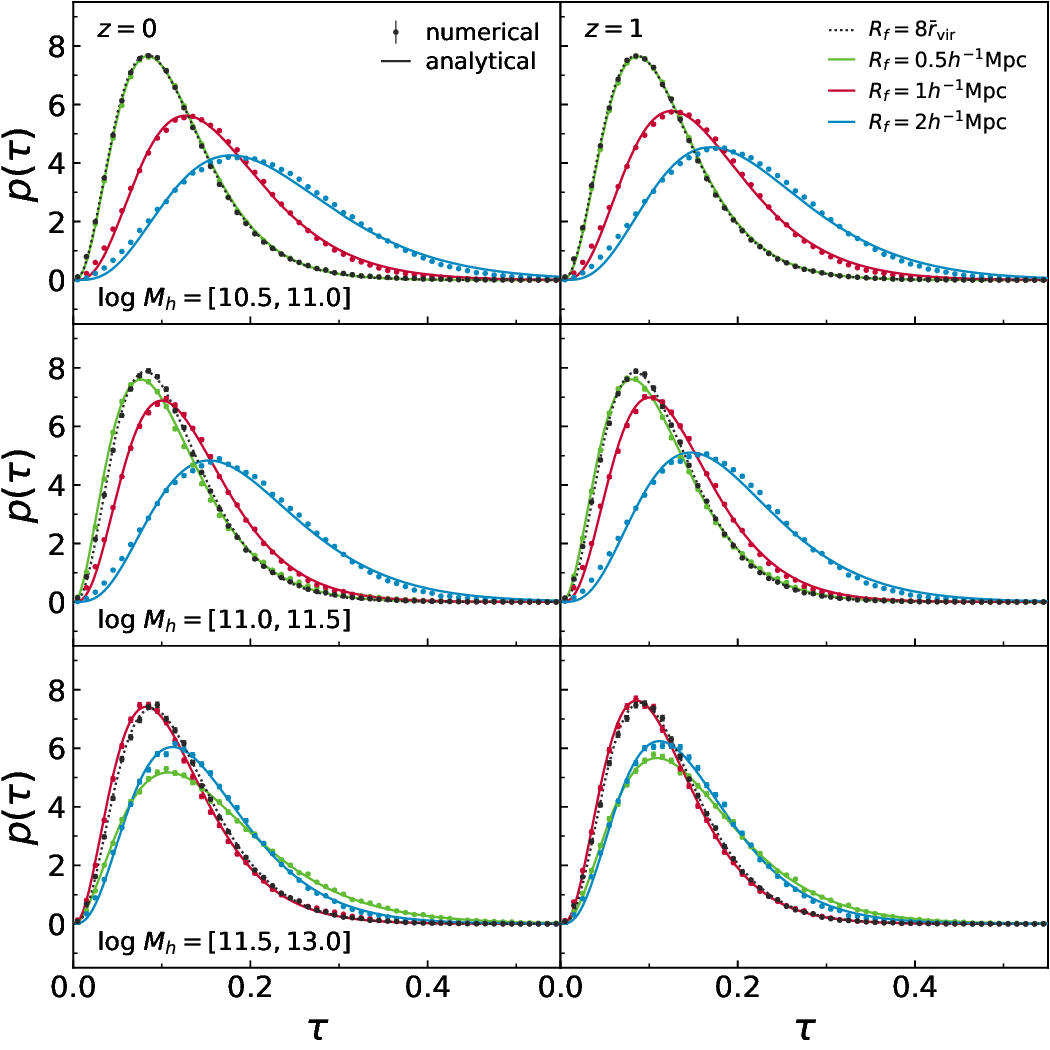}
\caption{Probability density functions of $\tau$ (filled red circles) with jackknife errors compared with the analytic $\Gamma$-distributions 
with the best-fit parameters (black solid lines) in three logarithmic mass ranges on four different scales at two different redshifts. 
Here, $\tau$ is the degree of the misalignments between the principal axes of protogalaxy inertia and initial tidal tensors at $z_{i}=127$, 
quantified by Equation (\ref{eqn:tau}). } 
\label{fig:pro_tau}
\end{figure}

\begin{figure}[ht]
\centering
\includegraphics[height=18cm,width=16cm]{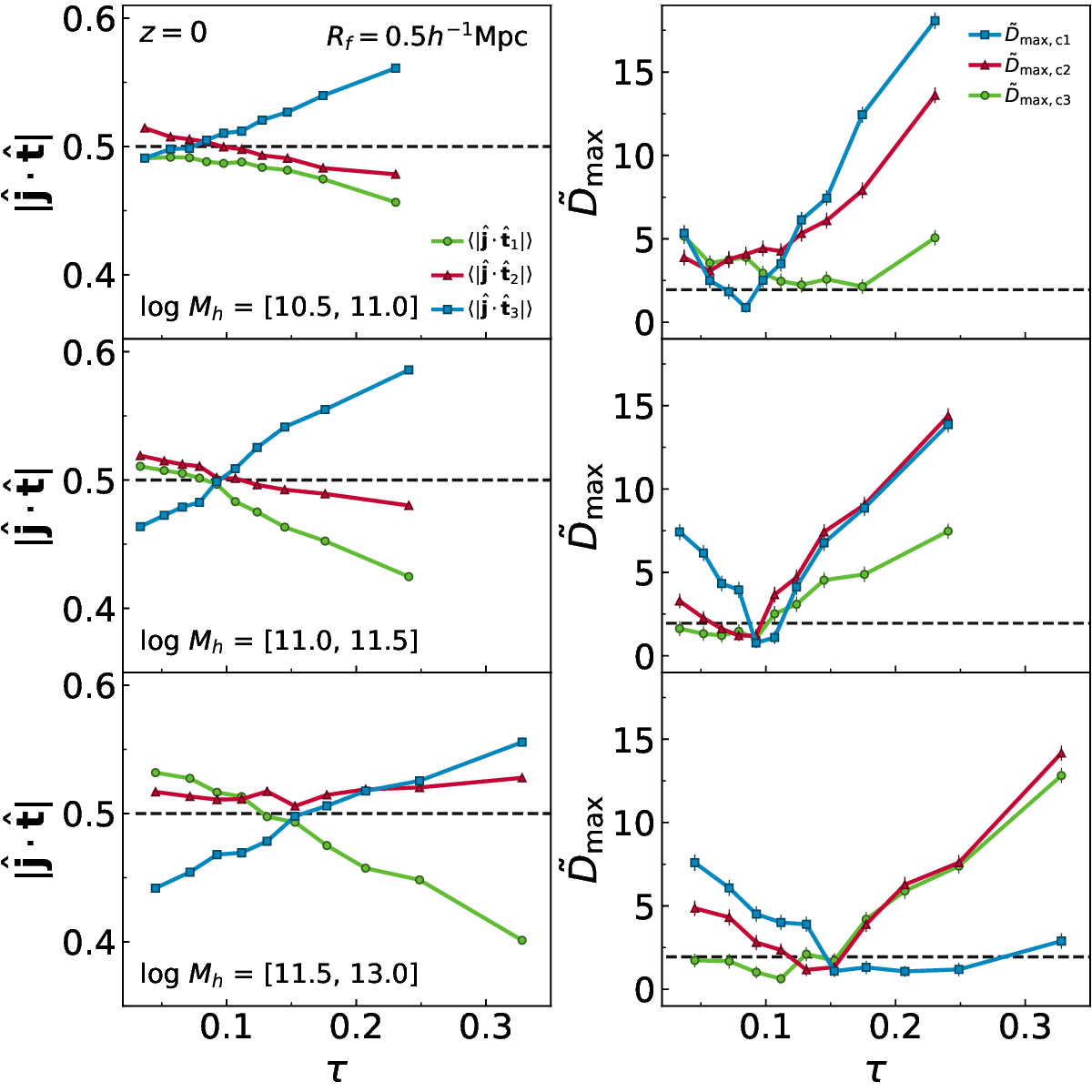}
\caption{(Left panels): Ensemble averages of the alignments between the galaxy spins and the major, intermediate 
and minor principal axes of the local tidal fields (filled green circles, red triangles and blue squares, respectively) in the three 
mass ranges on the scale of $R_{f}/(h^{-1}{\rm Mpc})=0.5$ at $z=0$. The horizontal dashed lines correspond to the case 
of complete misalignments. (Right panels): The confidence levels ({\it CL}) at which the KS test rejects  the null hypotheses of 
no occurrence of the types I, II and III spin transitions (filled blue squares, red triangles and green circles, respectively). 
The horizontal black dashed lines correspond to the $99.9\%$ {\it CL}. In each panel, the errors are estimated by the 
Jackknife method.}
\label{fig:align_z0_rf0.5}
\end{figure}

\begin{figure}[ht]
\centering
\includegraphics[height=18cm,width=16cm]{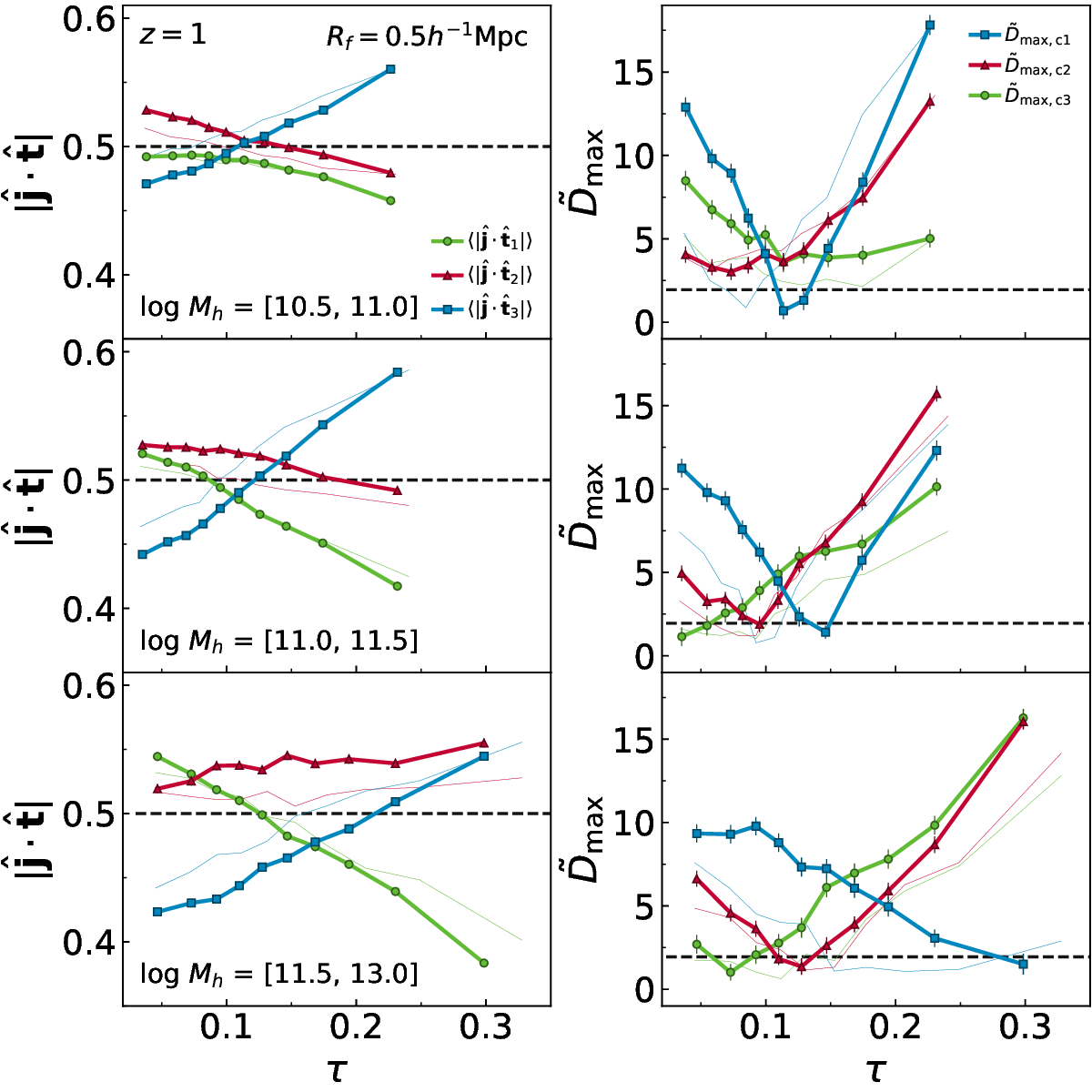}
\caption{Same as Figure \ref{fig:align_z0_rf0.5} but at $z=1$.  In each panel, the results for the case of $z=0$ and $R_{f}/(h^{-1}{\rm Mpc})=0.5$ 
are also shown as thin solid lines for comparison.}
\label{fig:align_z1_rf0.5}
\end{figure}

\begin{figure}[ht]
\centering
\includegraphics[height=18cm,width=16cm]{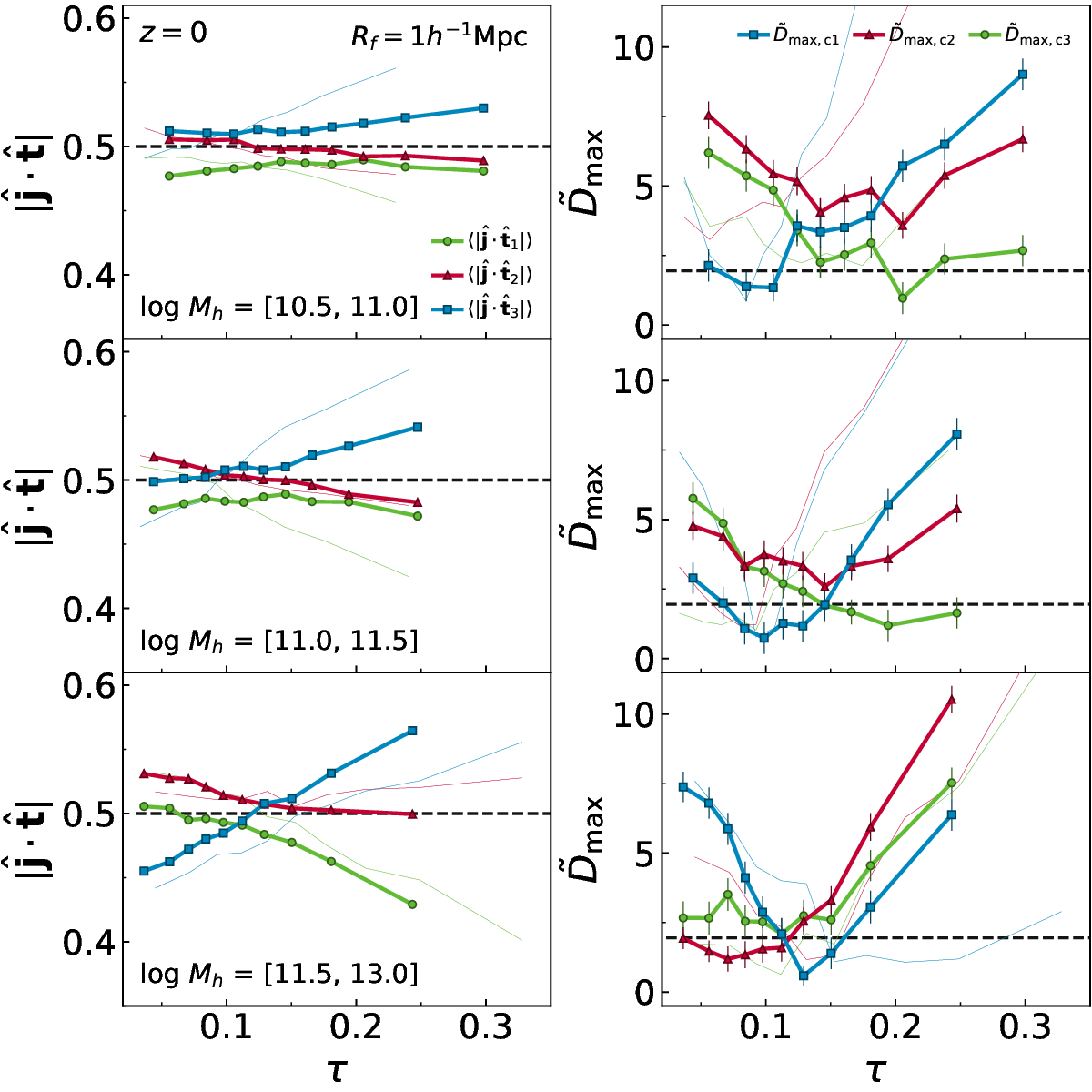}
\caption{Same as Figure \ref{fig:align_z1_rf0.5} but on the scale of $R_{f}/(h^{-1}{\rm Mpc})=1$ and $z=0$.}
\label{fig:align_z0_rf1}
\end{figure}

\begin{figure}[ht]
\centering
\includegraphics[height=18cm,width=16cm]{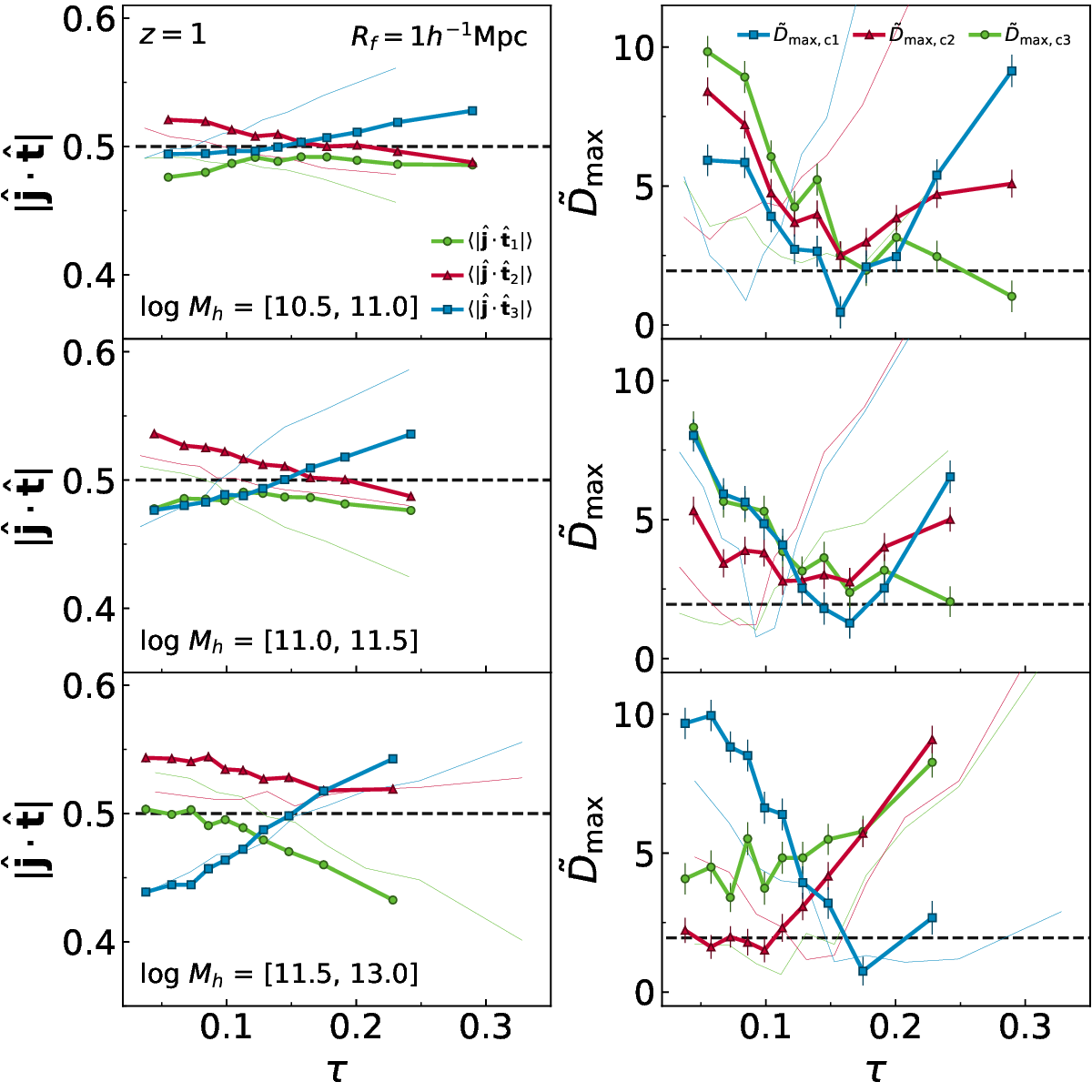}
\caption{Same as Figure \ref{fig:align_z1_rf0.5} but on the scale of $R_{f}/(h^{-1}{\rm Mpc})=1$.}
\label{fig:align_z1_rf1}
\end{figure}

\begin{figure}[ht]
\centering
\includegraphics[height=18cm,width=16cm]{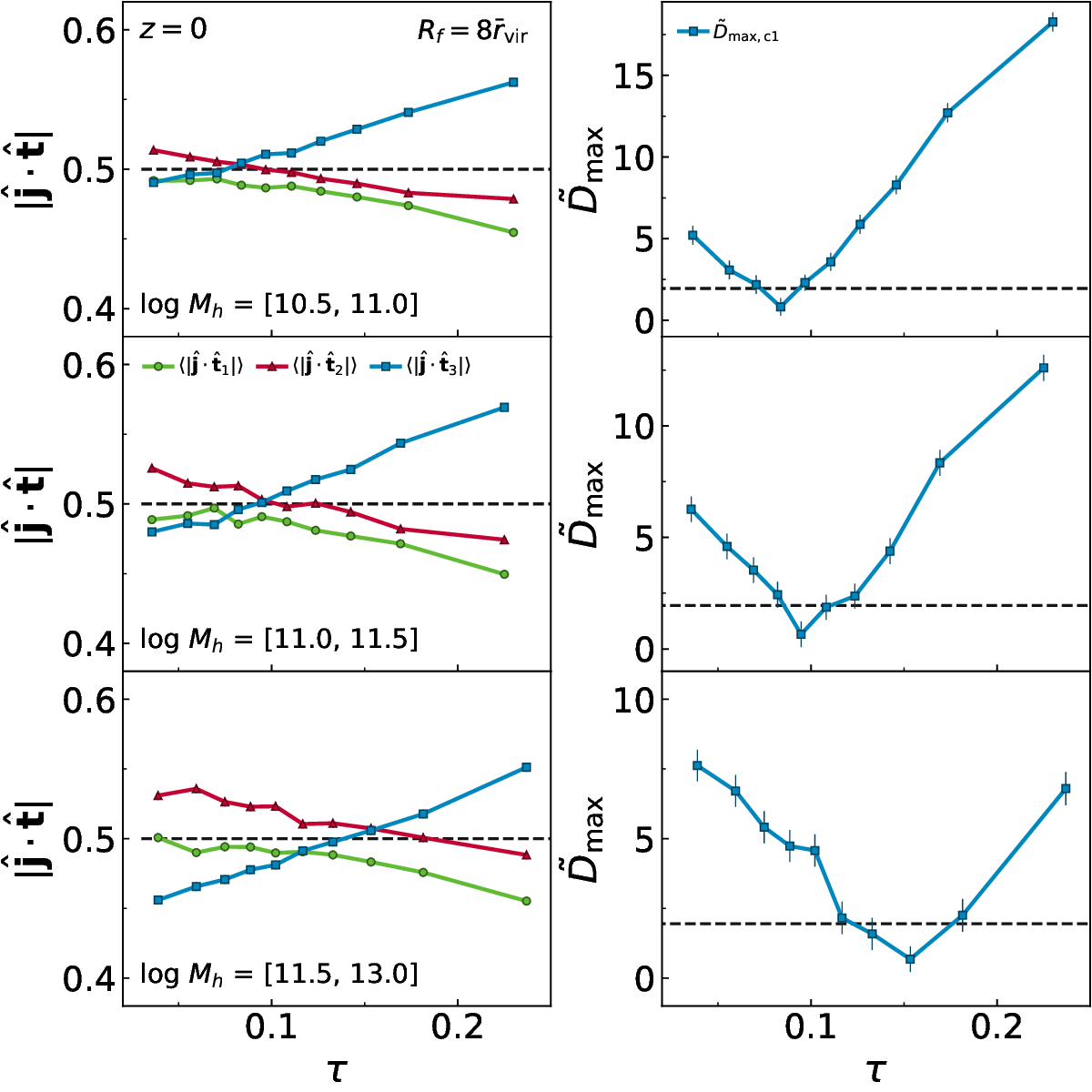}
\caption{Same as Figure \ref{fig:align_z0_rf1} but on the critical scale of $R_{f}/(h^{-1}{\rm Mpc})=8\bar{r}_{\rm vir}$ 
required for the universal form of $p(\tau)$.}
\label{fig:align_z0_rfc}
\end{figure}

\begin{figure}[ht]
\centering
\includegraphics[height=18cm,width=16cm]{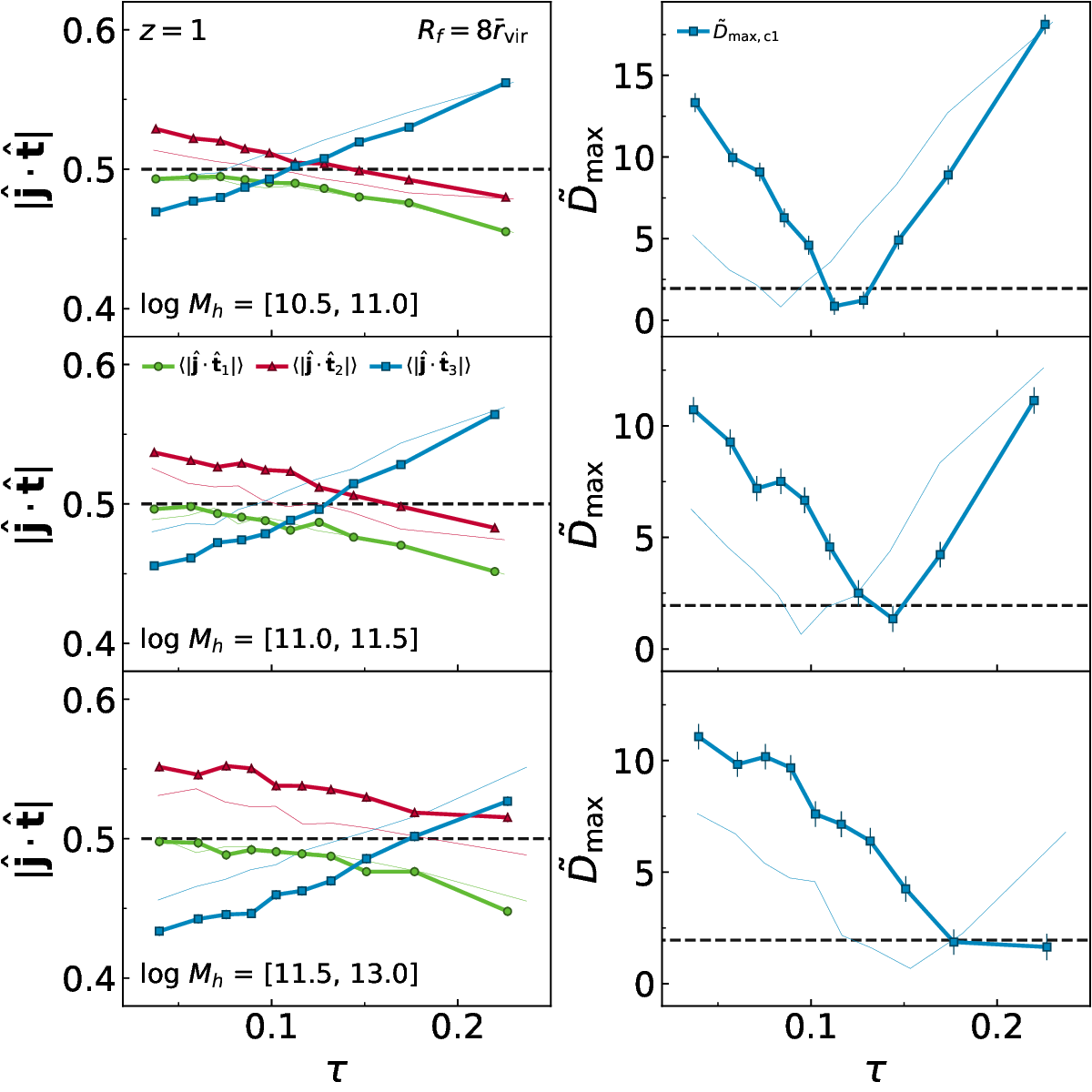}
\caption{Same as Figure \ref{fig:align_z0_rfc} but at $z=1$. In each panel, the results for the case of $z=0$ 
are also shown as thin solid lines for comparison.}
\label{fig:align_z1_rfc}
\end{figure}

\begin{enumerate}
\item
The analytic $\Gamma$-distribution characterized by two parameters, $k$ and $\theta$, excellently matches the numerically computed $p(\tau)$ for all of the cases 
(Figure \ref{fig:pro_tau}). 
When $r_{f}$ is set at a fixed value, the best-fit parameters vary with $m_{h}$ and $z$. However, when $r_{f}$ is set at a mass-dependent critical scale,  
$8\bar{r}_{\rm vir}$, the parameters become nearly robust, against the variation of $m_{h}$ and $z$, implying the universality of $p(\tau)$. 
\item
The type I transition of $\hat{\bf j}$ from $\hat{\bf t}_{3}$ to $\hat{\bf t}_{2}$ is witnessed as $\tau$ decreases below a threshold $\tau_{c1}$, 
for all of the considered cases of $m_{h}$, $z$ and $r_{f}$.   The transition signal is much stronger when $\langle\vert\hat{\bf j}\cdot\hat{\bf t}_{2}\vert\rangle$ and 
$\langle\vert\hat{\bf j}\cdot\hat{\bf t}_{3}\vert\rangle$ are expressed as a function of $\tau$ rather than of $m_{h}$, being consistently present in all of the three 
$m_{h}$-ranges. It implies that the previously found mass-dependent spin transition should be a mere reflection of the 
$\tau$-driven type I spin transition. 
\item
The type II transition of $\hat{\bf j}$ from $\perp\hat{\bf t}_{1}$ to $\parallel\hat{\bf t}_{1}$ is found to occur as $\tau$ decreases below a different threshold, 
$\tau_{c2}(<\tau_{c1})$. Unlike the type I,  the type II spin transition at $\tau_{c2}$ occurs only if $r_{f}<8\bar{r}_{\rm vir}$. 
Given the prior result that the protogalactic sites with higher $\tau$ tend to evolve into more rotationally supported late-type galaxies \citep{oh-etal13,jia-etal19,rod-etal22} 
with higher spin parameters, bluer colors, lower metallicities and younger ages \citep{ML23d}, it implies that the morphology-dependent spin transition found from the galaxies located 
on void surfaces \citep{LM23} is in fact a mere reflection of the $\tau$-driven type II spin transition. 
\item
In case that the inequality of $r_{f}\le 4r_{\rm vir}$ is satisfied, a strong signal of the type III transition of $\hat{\bf j}$ between $\hat{\bf t}_{2}$ and $\hat{\bf t}_{1}$ 
is detected as $\tau$ decreases below another threshold $\tau_{c3}(<\tau_{c2})$, even though $\hat{\bf j}$ is consistently measured  at $r_{\rm vir}$. 
It implies that the radius-dependent spin transition is in fact a mere reflection of the $\tau$-driven type III spin transition.
\end{enumerate}

The $\Gamma$-distribution is often used in statistics to model the mean waiting times between randomly occurring events \citep{hog-etal12}. Given our finding 
that $p(\tau)$ is excellently described by the $\Gamma$-distribution, we claim that $\tau$ may be a reaction time for a protogalaxy to withstand an external deforming effect 
and that $\tau$ plays the role of regulating a competition between the external tidal compression and the inner tension.  The more delayed the reaction time is, the more 
severely the tidal deforming effects become distracted, and thus the stronger the inner tension becomes. Although the density parity model proposed by 
\citet{ML23b} already suggested that this competition should be the key mechanism for the radius-dependent spin transition, two questions so far remained unanswered: 
what initial conditions control the competition and whether the other types could be explained by the same mechanism. 
The aforementioned speculation leads to the following interpretation of our new results, allowing us to find the answers to these two questions: 
\begin{enumerate}
\item 
In the lowest-$\tau$ section ($\tau<\tau_{c3}$) where ${\bf I}(z_{i})$ and ${\bf T}(z_{i})$ are almost perfectly anti-aligned with each other, the first order generation of the protogalaxy 
angular momentum is impossible, since ${I}_{12}={I}_{23}={I}_{31}=0$ in Equations (\ref{eqn:lttt1})--(\ref{eqn:lttt3}). 
Under this situation that there is no delay in the protogalaxy reaction, the effect of the tidal compression supersedes those of the linear tidal torque and protogalaxy inner tension, 
aligning $\hat{\bf j}$ with $\hat{\bf t}_{1}$ which is parallel to the minor principal axes of the inertia tensor. However, the tidal field can exert this dominant compression effect 
only on the scale of $r_{f}\le 4\bar{r}_{\rm vir}$. 
\item
In the medium-$\tau$ section ($\tau_{c3}\le \tau<\tau_{c1}$) where ${\bf I}(z_{i})$ and ${\bf T}(z_{i})$ are strongly but imperfectly anti-aligned with each other, 
the protogalaxy angular momentum can be generated at first order, as described by Equation (\ref{eqn:lp00}). 
The effect of the linear tidal torque supersedes those of the tidal compression and protogalaxy inner tension, aligning $\hat{\bf j}(z_i)$ with $\hat{\bf t}_{2}(z_i)$, 
as originally envisaged by \citet{lp00}.
\item
In the highest-$\tau$ section ($\tau\ge \tau_{c1}$) where ${\bf I}(z_{i})$ have appreciably large off-diagonal components, the approximation given in Equation (\ref{eqn:approx2}) 
is no longer guaranteed to be valid.  Under this situation, the protogalaxy reaction would be greatly delayed, the tidal compression becomes severely distracted especially along 
$\hat{\bf t}_{3}$ where the inner tension is most dominant, which results in the $\hat{\bf j}(z_i)$--$\hat{\bf t}_{3}(z_i)$ alignments.  
\item
Since the angular momentum is a conserved quantity, the galaxy spins at $z(\ll z_{i})$ retain the initially induced alignment tendencies, showing the $\tau$-driven transition 
behaviors with respect to the principal axes of the local tidal fields, depending on $r_{f}$. 
\end{enumerate}

Now that the galaxy spin transition has been found to be mainly driven by the initial condition $\tau$, it can also be naturally explained away why the spin transition thresholds 
depend sensitively on the background cosmology as found in the previous works \citep{lee-etal20,LL20}. It will be interesting to explore how differently the degree of 
misalignments between the protogalaxy inertia and initial tidal fields, $\tau$, are distributed in non-standard cosmologies, and whether or not the difference, if 
existent, can be used to probe the early universe physics. Our future works will be in this direction. 

\acknowledgments

The IllustrisTNG simulations were undertaken with compute time awarded by the Gauss Centre for Supercomputing (GCS) 
under GCS Large-Scale Projects GCS-ILLU and GCS-DWAR on the GCS share of the supercomputer Hazel Hen at the High 
Performance Computing Center Stuttgart (HLRS), as well as on the machines of the Max Planck Computing and Data Facility 
(MPCDF) in Garching, Germany.  
We thank an anonymous referee whose comments helped us improve the original manuscript. 
JSM acknowledges the support by the National Research Foundation (NRF) of Korea grant funded by the Korean government (MEST) (No. 2019R1A6A1A10073437).
JL acknowledges the support by Basic Science Research Program through the NRF of Korea funded by the Ministry of Education (No.2019R1A2C1083855).

\clearpage
\clearpage
\clearpage
\clearpage
\clearpage
\clearpage
\clearpage
\clearpage
\begin{deluxetable}{ccccccccc}
\tablewidth{0pt}
\tablecaption{Best-fit parameters of the $\Gamma$-distribution and three transition thresholds from $N_{g}$ TNG galaxies.}
\setlength{\tabcolsep}{2.3mm}
\tablehead{$z$ & $m_{h}$ & $r_{f}$  & $N_{g}$ & $k$ & $10^{2}\theta$ & $\tau_{c1}$ & $\tau_{c2}$ & $\tau_{c3}$}
\startdata
$0$ & $[10.5, 11)$ & $0.5$ &  $387262$ & $3.77 \pm 0.01$ & $3.04 \pm 0.01$ & $0.07$--$0.08$ & $-$ & $-$  \\
$0$ & $[11, 11.5)$ & $0.5$ &  $148145$ & $3.28 \pm 0.02$ & $3.35 \pm 0.02$ & $0.09$--$0.11$ & $0.07$--$0.09$ & $-$ \\
$0$ & $[11.5, 13)$ & $0.5$ &  $83491$ & $3.10 \pm 0.01$ & $5.11 \pm 0.02$ & $0.15$--$0.25$ & $0.13$--$0.15$ & $0.05$--$0.11$ \\
\hline
$0$ & $[10.5, 11)$ & $1$ &  $387262$ & $4.36 \pm 0.03$ & $3.78 \pm 0.03$ & $-$ & $-$ & $-$ \\
$0$ & $[11, 11.5)$ & $1$ &  $148145$ & $4.13 \pm 0.02$ & $3.19 \pm 0.02$ & $0.08$--$0.15$ & $-$ & $-$ \\
$0$ & $[11.5, 13)$ & $1$ &  $83491$ & $3.51 \pm 0.02$ & $3.28 \pm 0.02$ & $0.13$--$0.15$ & $0.04$--$0.11$ & $-$ \\
\hline
$0$ & $[10.5, 11)$ & $2$ &  $387262$ & $4.78 \pm 0.06$ & $4.71 \pm 0.07$ & $-$ & $-$ & $-$ \\
$0$ & $[11, 11.5)$ & $2$ &  $148145$ & $4.62 \pm 0.05$ & $4.24 \pm 0.06$ & $0.07$--$0.17$ & $-$ & $-$ \\
$0$ & $[11.5, 13)$ & $2$ &  $83491$ & $4.05 \pm 0.02$ & $3.68 \pm 0.02$ & $0.13$--$0.19$ & $-$ & $-$ \\
\hline
$1$ & $[10.5, 11)$ & $0.5$ &  $419169$ & $3.92 \pm 0.01$ & $2.97 \pm 0.01$ & $0.11$--$0.13$ & $-$ & $-$ \\
$1$ & $[11, 11.5)$ & $0.5$ &  $150370$ & $3.53 \pm 0.01$ & $3.19 \pm 0.01$ & $0.15$--$0.15$ & $0.10$--$0.10$ & $-$ \\
$1$ & $[11.5, 13)$ & $0.5$ &  $76121$ & $3.55 \pm 0.02$ & $4.27 \pm 0.02$ & $-$ & $0.11$--$0.13$ & $0.07$--$0.07$ \\
\hline
$1$ & $[10.5, 11)$ & $1$ &  $419169$ & $4.46 \pm 0.04$ & $3.63 \pm 0.03$ & $0.16$--$0.16$ & $-$ & $-$ \\
$1$ & $[11, 11.5)$ & $1$ &  $150370$ & $4.26 \pm 0.03$ & $3.08 \pm 0.02$ & $0.14$--$0.16$ & $-$ & $-$ \\
$1$ & $[11.5, 13)$ & $1$ &  $76121$ & $3.85 \pm 0.02$ & $3.01 \pm 0.01$ & $0.17$--$0.17$ & $0.09$--$0.10$ & $-$ \\
\hline
$1$ & $[10.5, 11)$ & $2$ &  $419169$ & $4.89 \pm 0.06$ & $4.36 \pm 0.06$ & $0.14$--$0.21$ & $-$ & $-$ \\
$1$ & $[11, 11.5)$ & $2$ &  $150370$ & $4.70 \pm 0.06$ & $3.97 \pm 0.05$ & $0.20$--$0.23$ & $-$ & $-$ \\
$1$ & $[11.5, 13)$ & $2$ &  $76121$ & $4.22 \pm 0.03$ & $3.47 \pm 0.03$ & $0.18$--$0.21$ & $-$ & $-$ \\
\enddata
\label{tab:bestfit1}
\end{deluxetable}

\clearpage
\begin{deluxetable}{ccccccccc}
\tablewidth{0pt}
\tablecaption{Same as Table \ref{tab:bestfit1} but on the critical scale of $R_{f}/(h^{-1}{\rm Mpc})=8\bar{r}_{\rm vir}$.}
\setlength{\tabcolsep}{3mm}
\tablehead{$z$ & $m_{h}$ & $r_{f}$  & $k$ & $10^{2}\theta$ & $\tau_{c1}$}
\startdata
$0$ & $[10.5, 11)$ & $0.485$ & $3.74 \pm 0.01$ & $3.04 \pm 0.01$ & $0.08$--$0.08$ \\
$0$ & $[11, 11.5)$ & $0.715$ & $3.75 \pm 0.02$ & $2.96 \pm 0.01$ & $0.09$--$0.11$ \\
$0$ & $[11.5, 13)$ & $1.253$ & $3.82 \pm 0.02$ & $3.13 \pm 0.01$ & $0.13$--$0.15$ \\
\hline
$1$ & $[10.5, 11)$ & $0.485$ & $3.89 \pm 0.01$ & $2.97 \pm 0.01$ & $0.11$--$0.13$ \\
$1$ & $[11, 11.5)$ & $0.713$ & $3.90 \pm 0.01$ & $2.90 \pm 0.01$ & $0.14$--$0.14$ \\
$1$ & $[11.5, 13)$ & $1.222$ & $4.07 \pm 0.02$ & $2.93 \pm 0.02$ & $0.18$--$0.23$ \\
\enddata
\label{tab:bestfit2}
\end{deluxetable}
\end{document}